\documentclass[twocolumn,showpacs,preprintnumbers,amsmath,amssymb]
{revtex4}

\usepackage{graphicx}
\usepackage{dcolumn}
\usepackage{bm}

\begin{document}

\preprint{APS/123-QED}

\title{Similarity of the 3.42 eV and near-band-edge 3.47 eV luminescence bands in GaN}

\author{T. V. Shubina}
\email{shubina@beam.ioffe.ru}
\author {S. V. Ivanov}
\author{V. N. Jmerik}
\author{D. D. Solnyshkov}
\author {N. A. Cherkashin}
\author{P. S. Kop'ev}
\affiliation{Ioffe Physico-Technical Institute, Polytekhnicheskaya
26, St.Petersburg 194021, Russia}

\author{A. Vasson and J. Leymarie}
\affiliation{LASMEA-UMR 6602 CNRS-UBP, 63177 AUBIERE CEDEX,
France}

\author{K. F. Karlsson}
\author{P. O. Holtz}
\author{B. Monemar}
\affiliation{Link\"oping University, S-581 83 Link\"oping, Sweden}

\date{\today}

\begin{abstract}
We demonstrate that the 3.42 eV photoluminescence (PL) band in GaN is of the same intrinsic origin as the near-edge $\sim$3.47 eV band, but arises from regions of inversed polarity characterized by different strain and growth rate. Two absorption edges are thermally detected at 0.35 K in nanocolumn structures, exhibiting both bands. Micro-PL studies have shown similar temperature/power behavior of these bands, with a competition in intensity in closely spaced spots accompanied by alterations of exciton level ordering. Strain-induced one-dimensional carrier confinement in small inversion domains likely explains the discrete narrow lines observed between the bands. 
\end{abstract} 


\maketitle
Despite thorough investigations of GaN, a wide-band gap semiconductor known as one of the most promising materials for highly efficient optoelectronics, some of its optical features are still under debate. Among them is the photoluminescence (PL) band detected at 3.41-3.42 eV (see Fig. 1 (a)), which lies significantly below the near-band-edge emission ($\sim$3.47 eV). The 3.42 eV band is observed in both GaN epilayers and nanocolumns \cite{1}-\cite{4}. The latter are of particular interest due to their possible application as nanocavities in polariton lasers and other vertical emitters \cite{5}. Such applications require a high optical quality of the nanocolumns that is incompatible with the appearance of the additional band. This is especially annoying because other properties of the nanocolumns may be excellent \cite{5},\cite{6}. 

In the previous studies, the origin of the 3.42 eV band has always been assumed to be extrinsic, associated with the excitons localized at deep donors or at dislocations \cite{1},\cite{2}. However, the band can appear or be absent in the structures grown in the same setup with a similar level of the background impurities, as well as in structures with the radically different dislocation density \cite{4}. 

In this paper, we claim the similarity of the 3.42 and 3.47 eV emission bands in GaN, assuming that both are intrinsic and originate from near-band-edge exciton states, but in spatially different regions, generally having opposite polarities. In other words, inversion domains (IDs) are responsible for splitting of the photoluminescence band in GaN.

\begin{figure}
\includegraphics{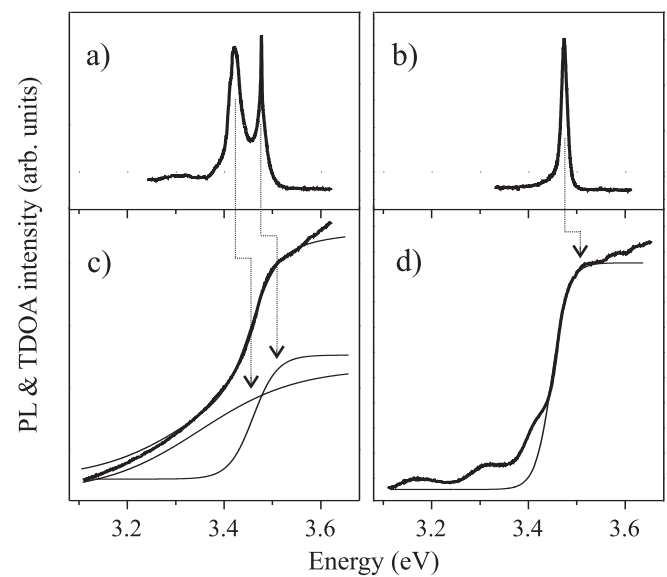}
\caption{\label{fig:epsart} 
Surface $\mu-$PL (a, b) and TDOA (c, d) spectra taken in an ID-enriched nanocolumn sample and in a reference GaN epilayer, respectively. The periodic structure below the principal edge in (d) is due to interference in the perfect GaN layer. Arrows indicate band edges as determined by the TDOA technique. Thin lines are fittings of the absorption spectra. 
}
\end{figure}

The IDs are prevailing structural defects in GaN, characterized by an opposite sequence of Ga and N atoms compared to the surrounding area. In general opinion, this is the only difference. However, our recent study of GaN quantum wells has shown that the emission in the ID regions is strongly shifted to lower energies, in part due to different strain \cite{7}. This fact is consistent with the generally recognized dependence of transition energies on strain in GaN \cite{8}. We assume that the strain factor plays essential role in any structures containing IDs. 

In the study we focus on the GaN nanocolumns. These separately standing objects are most suitable for optical investigation with high spatial resolution, e.g. by a micro-PL ($\mu-$PL) technique. The nanocolumns were grown on c-sapphire at the substrate temperature T$_S$=700-750$^\circ$C by plasma-assisted molecular beam epitaxy \cite{6}. The growth procedure included the substrate heating and nitridization at T$_S$=800$^\circ$C and 750$^\circ$C, respectively. The samples were grown either directly on the nitridized substrate or on a special 20-30 nm-thick low-temperature (T$_S$=300$^\circ$C) buffer. 

A transmission electron microscopy (TEM) study of the samples has revealed many IDs (Fig. 2). The inversion of the polarity is confirmed by careful examination of an interface region which contains a set of parallel dark and bright lines related to the (0001) stacking faults. The change of the upper line contrast corresponds to the flipping of the polarity. The IDs are usually located between the nanocolumns, in their joints. A preliminary micro-cathodoluminescence study demonstrated that the 3.42 eV band is most intense in these regions, being stronger at a certain depth \cite{4}. This is consistent with the finding that the IDs stop growing before the nanocolumns do. 

\begin{figure}
\includegraphics{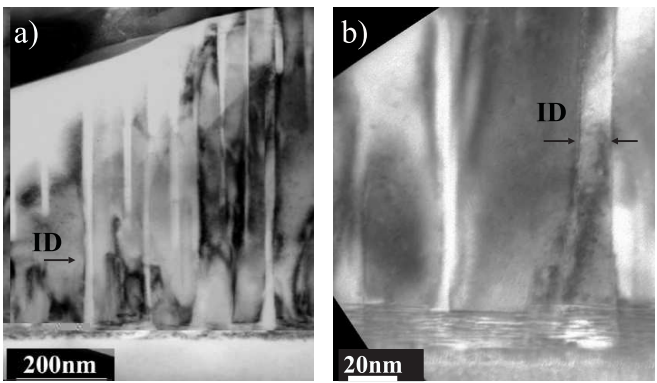}
\caption{\label{fig:epsart}
Dark field cross-section TEM images of a nanocolumn sample enriched by IDs (a) with B=[2-1-10], g=$<01-10>$; (b) B=[2-1-10], g=$<0002>$.}
\end{figure}

Two absorption edges have been found in this ID-enriched sample. This result has been obtained using an original technique of thermally detected optical absorption (TDOA). The method allows one to study thick or inhomogeneous layers, like the separately standing nanocolumns, where the conventional optical absorption technique fails. These measurements, performed at 0.35 K, are based on the detection of a small increase in the sample temperature. This increase arises from the creation of phonons via nonradiative recombination activated by the initial optical absorption \cite{9}. 

Our studies of reference GaN epilayers have shown that the absorption edge in a TDOA spectrum corresponds to the kink between flat and slope parts. For instance, the kink in the spectrum of a perfect GaN epilayer is at 3.50 eV (Fig. 1 (d)), which coincides very well with the bandgap of almost unstrained GaN \cite{10}. The kink energy is $\sim$30 meV higher than the maximum of the main PL line, attributed usually to the donor bound exciton. This binding energy is approximately equal to a sum of binding energies of free and donor bound excitons in GaN \cite{11},\cite{12}. The absorption tail in the GaN epilayer satisfies formally the Urbach-Martienssen rule $\alpha=\alpha_{0}exp((h\nu-E_{0})/\mu)$, where $E_{0}$, $\alpha_{0}$ are characteristic material parameters, $\mu=k_{B}T/\delta$ depends on a measure of exciton-phonon coupling $\delta$ in a material at a certain temperature $T$, $k_{B}$ is the Boltzmann constant \cite{13}. A good fit of the spectra in Fig. 1 (d) has been obtained at $\mu\sim0.018$ eV.

The TDOA spectra registered on the nanocolumn samples exhibit an additional tail extended to the lower energy. The tail is so pronounced that these spectra can be fitted using two absorption edges (Fig. 1 (b)). The higher-energy edge is near the same 3.50 eV, while the other is markedly lower, near 3.45 eV. It is worth noting that commonly known defects cannot provide such a strong absorption below the principal absorption edge, while differently strained local regions can do that. The parameter $\mu$ which determines the steepness of the tail, is significantly higher for the 3.45 eV edge than for the 3.50 eV one (0.1 vs 0.025 eV). Here, the broadening is not only due to the phonon interaction, but results from a superposition of absorption in ID regions which may be under different strain. 

\begin{figure}
\includegraphics{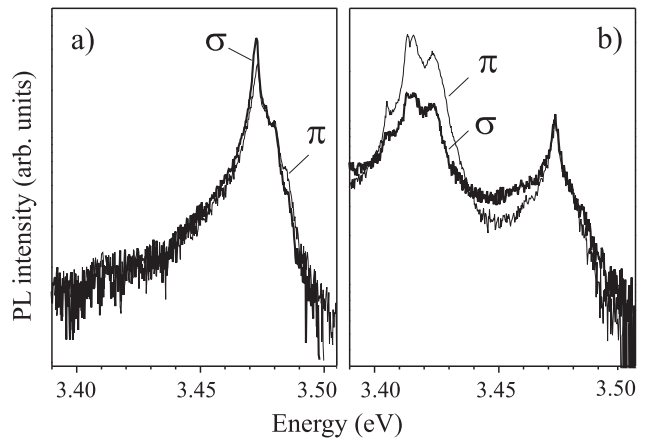}
\caption{\label{fig:epsart} 
Polarized $\mu$-PL spectra taken at 2 kW/cm$^2$ in two closely-spaced cleaved-edge spots in a nanocolumn sample with dominant (a) 3,47 eV and (b) 3.42 eV bands. 
}
\end{figure}

We have studied $\mu$-PL from both surface and cleaved-edge of the sample. The $\mu$-PL technique has spectral and spatial resolutions of about 0.5 meV and 1.5 $\mu$m, respectively. The measurements were carried out in a He continuous flow cryostat under cw excitation by a 266 nm laser line. With the cleaved-edge measurements, the vector $\bold{k}$ of the exciting light is normal to the main nanocolumn axis $\bold{c}$, and a signal in two linear polarizations - $\sigma$ ($\bold{E} \perp \bold{c}$) and $\pi$ ($\bold{E} \| \bold{c}$) can be registered. 

It has been found that the two-band emission with characteristic energies of 3.47 and 3.42 eV varies strongly between neighboring spots (Fig. 3). When the 3.42 eV emission is absent, the near-band-gap PL is strong and exhibits well-pronounced free-exciton features, whose relative intensity and energy position satisfy the selection rules of a wurtzite crystal, similar to previously observed in $\mu-$PL spectra of thick GaN layers \cite{14}. The appearance of the 3.42 eV lines is accompanied by a decrease of the 3.47 eV PL intensity and suppression of the free excitons. 

The most intricate finding is the change of dominant polarization along the spectra - the 3.42 eV band turns out to be predominantly $\pi$-polarized, while the near-edge emission is $\sigma$-polarized. The change of the polarization occurs in the vicinity of 3.45 eV. One can suppose that we have observed the strain-induced change of the $\Gamma_9$ and $\Gamma_7$ exciton ordering, which takes place at the tensile strain $\epsilon_z$ of about -0.001 in the vicinity of 3.45 eV \cite{10}. Transitions involving carriers localized at the ID boundaries have also been predicted to be $\pi$-polarized \cite{15}. However, this prediction cannot explain a large shift of the emission relatively to the main band ($\sim$ 50 meV). 

Power and temperature dependences of both bands are similar (Fig. 4). There are sharp ($\sim$3 meV wide) peaks in both bands at low excitation power densities. With a power rise, broader peaks appear in both bands. Such peaks in the near-band-edge 3.47 eV emission are associated with the donor bound and free excitons, respectively. We believe that similar peaks in the 3.42 eV band have the same origin. With a temperature rise, the higher-energy part of both bands is relatively increased, while the lower-energy narrow peaks decay rapidly. Such a temperature behavior is characteristic for free and bound exciton emission, respectively \cite{11}. A partially resolved structure of the 3.42 eV PL had been observed previously as well \cite{3}. The true interpretation was likely hampered by the presence of a set of lines from closely spaced regions. 

\begin{figure}
\includegraphics{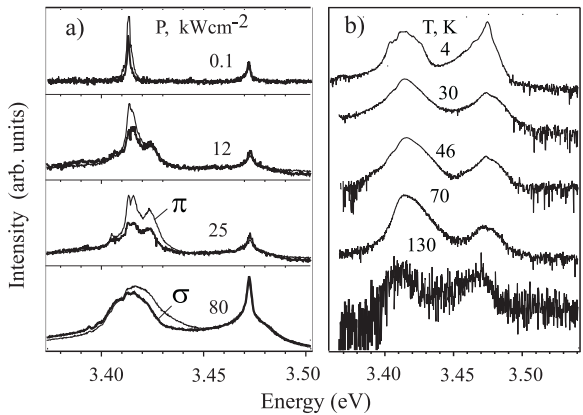}
\caption{\label{fig:epsart} 
Cleaved-edge $\mu$-PL spectra of the nanocolumn sample: (a) $\pi$- and $\sigma$-polarized components at different excitation power densities; (b) variation of non-polarized emission at a temperature rise. 
}
\end{figure}

\begin{figure}
\includegraphics{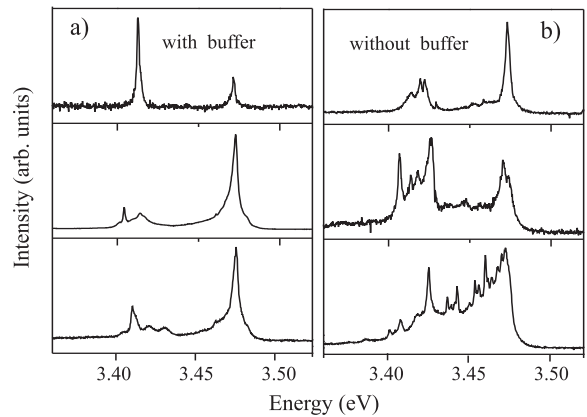}
\caption{\label{fig:epsart} 
Variation of cleaved-edge $\mu-$PL spectra between closely-spaced spots in nanocolumn samples grown (a) with (b) without the low-temperature buffer. 
}
\end{figure}

The 3.42 eV band is not homogeneous, but often consists of discrete narrow peaks (Fig. 5). Moreover, the narrow peaks can appear between these two principal bands. The number of these peaks and their energy position are dependent on the existence of a buffer layer, its quality and fabrication technique. We have registered narrow lines with spectral width as narrow as 1.5 meV at energies significantly higher than 3.42 eV in the structures grown without any buffer, which are characterized by many small IDs.

It has been reported long ago that strain-induced confinement can form a one-dimensional density of states in quantum-wires (QWRs). This was achieved, e.g., by modulation of the in-plane lattice constant of QWs by means of a complicated growth procedure \cite{16}. In the nanocolumn structure without any buffer, the small IDs are formed predominantly. These small IDs, being under different strains compared with adjacent regions, can be considered as such QWRs formed in a natural way. The transition energies in the QWRs are shifted up by the quantum confinement effect with respect to 3.42 eV. The shift depends on both the ID characteristic diameter and the strain. Variation of the parameters can determine the observed variety of the narrow-line energies. 

To study peculiarities of growth with different strains in adjacent regions we have performed the dynamic Monte-Carlo simulations. We have extended the model suggested by Plotz et al \cite{17}. To speed up the calculations, only a two-dimensional cross-section along the (100) plane of quasi-cubic structure possessing equilibrium lattice constant $r_0$ has been considered. Atomic interaction was described by the directional Lennard-Jones potential:
\begin{equation}
E_{12}(r,\phi_1,\phi_2)=
E_0\frac{r^{-12}-(1-\beta)\alpha
r^{-6}-\beta f(\phi_1)f(\phi_2)\alpha r^{-6}}{\alpha^2}
\nonumber
\end{equation}
$$
f(\phi_{1,2})=cos^2(2\phi_{1,2}),  \alpha=2r_0^{1/6}
$$
The parameters of this potential for GaN are taken as follows: binding energy $E_0$=2.24 eV, degree of covalence $\beta=0.8$ \cite{18}. At long distances ($r>2.5r_0$) the potential was assumed to be zero. Compressive and tensile strains were modeled by means of decreasing or increasing distance between vertical atomic planes in corresponding part of the system. The difference between $r_0$ and the lattice constant of the strained area is taken as 5\% for the sake of demonstration. The atoms were successively deployed on a surface consisting of two different regions (strained and strain-free). The position for a new atom was chosen at random, but it had to decrease system energy to guarantee a certain bond between the atom and the surface. Note that this simulation cannot elucidate how the strain difference is created during initial growth stages.

\begin{figure}
\includegraphics{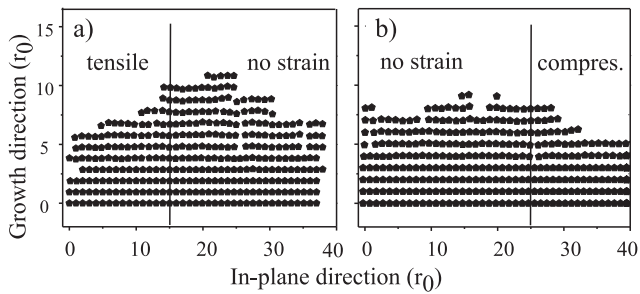}
\caption{\label{fig:epsart} 
Monte-Carlo simulation of epitaxial growth at different strain in adjacent regions (a) compressive and zero strain (b) tensile and zero strain.
}
\end{figure}

The most essential result of the simulation is presented in Fig. 6. One can see that any strain, tensile or compressive, causes a decrease of the growth rate. The IDs in the nanocolumns are most likely under tensile strain according to our optical study. This tensile strain inhibiting the growth can obviously be one of the factors facilitating the formation of the well-separated columns. These findings are well consistent with previously observed different growth rate in the regions of opposite polarities \cite{19}, as well as with recognition of a role of the IDs in the nanocolumn formation \cite{20}. 

In conclusion, the 3.42 and 3.47 eV bands of emission from GaN are analogous with each other. Both are near-band-edge emission split by different strain in regions of opposite polarities. This 3.42 eV band is not homogeneous; its fine structure depending on growth conditions may be characteristic either for "bulk" GaN under tensile strain or strain-induced quantum wires.

We thank Prof. A. Kavokin for fruitful discussion. This work was partly supported by RFBR and the cooperative Russia/France Grant N04509PB.



\end{document}